**Title:** Revisiting resource selection probability functions and single-visit methods: Clarification and extensions

**Authors:** Péter Sólymos[1,2,3,*], Subhash R. Lele[4,5]

[1] Alberta Biodiversity Monitoring Institute and Boreal Avian Modelling project, Department of Biological Sciences, CW 405, Biological Sciences Building, University of Alberta, Edmonton, Alberta T6G 2E9 Canada

[2] Boreal Avian Modelling Project, 751 General Services Building, University of Alberta, Edmonton, Alberta T6G 2H1 Canada

[3] Department of Biological Sciences, University of Alberta, Edmonton, Alberta T6G 2E9 Canada

[4] Department of Mathematical and Statistical Sciences, University of Alberta, Edmonton, Alberta T6G 2G1 Canada

[5] Department of Mathematical Sciences, Centre for Biodiversity Dynamics, Norwegian University for Science and Technology, Realfagsbygget, NO-7491 Trondheim, Norway

* Corresponding author: Péter Sólymos, Alberta Biodiversity Monitoring Institute and Boreal Avian Modelling Project, Department of Biological Sciences, CW 405, Biological Sciences Building, University of Alberta, Edmonton Alberta T6G 2E9 Canada, fax: 780-492-7635, e-mail: Sólymos@ualberta.ca

**Running head:** Revisiting single-visit methods

**Word count**: 6630 from Introduction to Figure legends

**Article type:** Forum article

**Summary**

(1) Models accounting for imperfect detection are important. Single-visit methods have been proposed as an alternative to multiple-visits methods to relax the assumption of closed population. Knape and Korner-Nievergelt (2015) showed that under certain models of probability of detection single-visit methods are statistically non-identifiable leading to biased population estimates.

(2) There is a close relationship between estimation of the resource selection probability function (RSPF) using weighted distributions and single-visit methods for occupancy and abundance estimation. We explain the precise mathematical conditions needed for RSPF estimation as stated in Lele and Keim (2006). The identical conditions, that remained unstated in our papers on single-visit methodology, are needed for single-visit methodology to work. We show that the class of admissible models is quite broad and does not excessively restrict the application of the RSPF or the single-visit methodology.

(3) To complement the work by Knape and Korner-Nievergelt, we study the performance of multiple-visit methods under the scaled logistic detection function and a much wider set of situations. In general, under the scaled logistic detection function multiple-visits methods also lead to biased estimates.

(4) As a solution to this problem, we extend the single-visit methodology to a class of models that allows use of scaled probability function. We propose a Multinomial extension of single visit methodology that can be used to check whether the detection function satisfies the RSPF condition or not. Furthermore, we show that if the scaling factor depends on covariates, then it can also be estimated.

(5) We argue that the instances where the RSPF condition is not satisfied are rare in practice. Hence we disagree with the implication in Knape and Korner-Nievergelt (2015) that the need for RSPF condition makes single-visit methodology irrelevant in practice.

**Introduction**

Occupancy models (MacKenzie et al. 2002) and N-mixture models (Royle 2004) are popular approaches to deal with imperfect detection of unmarked organisms. These methods, in their original formulations, require replicate-visits to sites. For the identifiability of the model parameters, these methods need that the population be closed during the replicate-visits, i.e. there is no emigration, immigration, births or deaths between the visits. This assumption is often difficult to satisfy in practice and can lead to biased estimates of occupancy or abundance (Rota et al. 2009, Bayne et al. 2011). Logistical and cost related issues arise naturally when conducting repeated surveys. For example, is it pragmatic to visit many sites a small number of times or a small number of sites many times? Lele et al. (2012) and Sólymos et al. (2012) proposed a solution where they showed that under some easily satisfied conditions the single-visit surveys could be used to correct for imperfect detection. This avoids the need for the closed population assumption and eliminates the costs associated with repeated visits. The single-visit approach is logistically simpler and can be applied to historical data sets that lack replicated surveys. It can also be argued that, with the same budget, a single-visit methodology allows researchers to study larger geographical areas than replicate-visit methods and thus gain greater generality. Given the financial limitations most ecological studies face, this can be an important deciding factor.

In a recent paper, Knape and Korner-Nievergelt (2015), henceforth referred to as K&K, showed that under the log-link and the scaled logistic model for probability of detection, the single-visit method can estimate relative change in the abundances but not the absolute abundances. They also point out the relationship between their results and possible problems in the estimation of resource selection probability functions (RSPF) as suggested in Lele and Keim (2006). We are very grateful to the authors for pointing out these issues and particularly the relationship with results in Lele and Keim (2006).

The comments by K&K have made us think about our methodology in a rigorous fashion. We realized that in our papers on single-visit methodology we should have stated the necessary conditions on the class of permissible models for probability of detection and occupancy under which the identifiability holds. Such a condition, which we now refer to as RSPF condition, was stated in Lele and Keim (2006). If the RSPF condition is satisfied, it is possible to estimate absolute probability of selection. Similarly if the models for probability of detection satisfy the RSPF condition, the results related to single-visit methodology remain valid. We note that both K&K and Hastie and Fithian (2013) observe non-identifiability because the log-link and the scaled logistic model do not satisfy the RSPF condition. Fortunately, neither the log-link nor the scaled logistic model has found much use in practice. In our experience, most models that are actually used in practice do satisfy the RSPF condition. In this paper, we argue that providing recommendations for statistical methods should not ignore practical considerations, namely, how often an extreme situation when the method fails is likely to happen *in practice*.

**The RSPF condition: permissible class of models**

*Estimation of the probability of selection*

For pedagogical reasons, we start with the problem of estimation of absolute probability of selection from the use-available data (Lele and Keim 2006). Let $f^A(\underline{x})$ and $f^U(\underline{x})$ denote the distribution of resources on available and used units respectively. They are related to each other as: $f^U(\underline{x}) = \frac{\pi(\underline{x};\beta)f^A(\underline{x})}{\int \pi(\underline{x};\beta)f^A(\underline{x})d\underline{x}}$ where $\pi(\underline{x};\beta)$ denotes the probability of selecting the resource $\underline{x}$, given that it is encountered (Lele and Keim 2006; Lele et al. 2013). It is called the resource selection probability function (RSPF). By definition, the function $\pi(\underline{x};\beta)$ is any function that takes values between 0 and 1. The ratio function $\pi(\underline{x};\beta)/\pi(\underline{y};\beta)$, the relative probability of selection, is commonly known as the Resource Selection Function (RSF). We emphasize that this ratio function need not be an exponential function. For example, if $\pi(\underline{x};\beta)$ is based on a logit or complementary log-log link, the ratio function $\pi(\underline{x};\beta)/\pi(\underline{y};\beta)$ is not an exponential function.

It is well established that, given data arising from $f^U(\underline{x})$, one can estimate the ratio $\pi(\underline{x};\beta)/\pi(\underline{y};\beta)$. However, because the scientific interest lies in estimating $\pi(\underline{x};\beta)$, Lele and Keim (2006) asked the question: If we have an estimate of the ratio $\pi(\underline{x};\beta)/\pi(\underline{y};\beta)$, what conditions will allow inference on $\pi(\underline{x};\beta)$? Or, equivalently, can we identify the parameter $\beta$ using only the knowledge of the ratio function? The necessary condition for this was found to be as follows: The RSPF $\pi(x;\beta)$ should be such that if $\beta \neq \beta^*$, then $\sup_x |\pi(x;\beta) - K\pi(x;\beta^*)| > 0$ for any constant $K > 0$ (Lele and Keim, 2006; Gilbert et al. 1999). For ease of reference, we call this the RSPF condition and the class of models that satisfy this condition as RSPF model class. Loosely speaking, the RSPF condition reduces to the following two conditions: (1) Not all covariates in the model are categorical, and (2) the function $\log \pi(\underline{x};\beta)$ is non-linear and involves all components of the parameter vector.

We first note that the exponential function or scaled logistic (or, in general any scaled function) with unknown scaling constant, do not satisfy the RSPF condition. If all covariates in the model are categorical covariates with finite number of categories, from the log-linear model theory it follows that the probability of selection is necessarily modeled by the exponential function. Hence we need the condition that not all covariates are categorical. In the following, we explain the mathematical reasoning behind the RSPF condition in a simple situation.

A necessary step in any model fitting exercise is that we have an idea about the form of the model that we want to fit. Hence we assume that the model form $\pi(\underline{x}; \beta)$ is specified. Given this model specification, the goal of statistical analysis is to infer about the parameter $\beta$. First we note that if we know the ratio $\pi(\underline{x}; \beta)/\pi(\underline{y}; \beta)$, we know the difference $\log \pi(\underline{x}; \beta) - \log \pi(\underline{y}; \beta)$. For the simplicity of the argument, let us consider the case where there is only one continuous covariate and that the function is differentiable in $x$. Then, knowing the difference $\log \pi(\underline{x}; \beta) - \log \pi(\underline{y}; \beta)$ for every pair $(x,y)$ is equivalent to knowing $\frac{d}{dx} \log \pi(x; \beta)$. It follows that only those parameters that are involved in the function $\frac{d}{dx} \log \pi(x; \beta)$ are potentially identifiable. Further, the RSPF condition can be restated as a necessary and sufficient condition: $\frac{d}{dx} \log \pi(x; \beta) = \frac{d}{dx} \log \pi(x; \beta^*)$ if and only if $\beta = \beta^*$. The RSPF condition restricts the model space for $\pi(x; \beta)$ so that $\frac{d}{dx} \log \pi(x; \beta)$ depends on all components of the parameter vector $\beta$. This leads to a simple conclusion that RSPF model class consists of functions that can be written as: $\pi(x; \beta) = c(\beta)g(x; \beta)$ where $c(\beta)$ is not a constant function of $\beta$. Simple calculus shows that if RSPF is based on logit or complementary log-log link function, the RSPF condition is satisfied. Consideration of the polynomial functions in the exponent part of the logistic or complementary log-log functions provides us with a very flexible class of models that satisfy the RSPF condition (Figure 1). Such functions take values between 0 and 1 but may never reach or even come arbitrarily close to either of the boundary values (Figures 1 and 2). In the Supporting Information we provide a simple program to generate data under any RSPF model and to estimate its parameters. Readers can try different model forms for $\pi(x;\beta)$ and see for themselves if the methods work or not. Remember that if the non-linearity on the log-scale is weak, one may need very large sample sizes to get reasonable estimates.

*Single-visit occupancy studies*

Let *z* and *x* denote the covariates that affect detection and occupancy respectively. Let $p(z;\theta)$ and $\Psi(x;\beta)$ denote the probability of detection and probability of occupancy respectively. Again, for the sake of simplicity, we will consider single continuous covariates. It is trivial to see that we can estimate $p(z;\theta)\Psi(x;\beta)$ from single-visit data. The question is: Given this product, when can we separate out the components?

For notational simplicity, let us denote the product function $p(z;\theta)\Psi(x;\beta)$ by $h(x,z;\beta,\theta)$. We first note that if we know $h(x,z;\beta,\theta)$ and one of the components, say $p(z;\theta)$, then we can obtain the other component, $\Psi(x;\beta)$. Second thing to note is that the ratio $p(z;\theta)/p(z';\theta)$ can be obtained from the product function by $h(x,z;\beta,\theta)/h(x,z';\beta,\theta)$. Hence it follows that the necessary condition for identifiability is that $p(z;\theta)$ belongs to the RSPF model class. Conversely, we may impose the condition that $\Psi(x;\beta)$ belongs to the RSPF model class. The RSPF condition needs to be satisfied by at least one of the two components. This allows us to decompose the product function into two components. However, to determine which component is detection and which component is occupancy, we need to impose the condition that the set of covariates that affect detection and covariates that affect occupancy should not be completely overlapping.

Because of the low information content in binary data, estimation of occupancy and detection from single-survey data is extremely difficult (Welsh et al. 2013). As described in Lele et al. (2012) and Moreno and Lele (2010), one may need to use penalized likelihood function to stabilize the estimators. Unfortunately it is not clear how to choose a good penalty function in general. In the supplementary information, we use a quasi-Bayesian approach where the means of the prior distributions for the parameters are determined from the observations themselves instead of based on a probabilistic quantification of 'belief'. This seems to stabilize the estimation process considerably, resulting in estimators that are nearly unbiased. Unfortunately this estimation method lacks a strong theoretical basis. We are currently exploring the use of expert opinion (Lele and Allen 2006) to stabilize the estimators in this situation.

*Single-visit abundance surveys*

As compared to species occupancy, abundance data are more informative and hence there seems to be little need for stabilizing the estimators. Under the log-link model for the abundances, it is clear that one can obtain estimates of $p(z;\theta)\exp(X\beta)$. Hence, following the logic described above, as long as $p(z;\theta)$ belongs to the RSPF model class, we can estimate the abundances using single-visit survey data, irrespective of the link function for the abundance.

Notice that RSPF model class does not require models to reach the boundary values 0 or 1 for any covariate combination (Figure 2). In the supplementary material, we provide programs that the readers can use to test the results described above. We also note that, given the non-linear nature of the problem, it is analytically impossible to know if the solution to the likelihood equation will be unique. One may use the data cloning method (Lele et al. 2010, Sólymos 2010, Campbell and Lele 2014) to diagnose the non-uniqueness of the solution.

**Sensitivity to model assumptions**

We have now clarified the precise mathematical conditions under which single-survey methodology is valid. Clearly the log-link and scaled logistic model used in K&K do not satisfy the RSPF condition and hence if the true detection probability function is a scaled logistic function, the estimates are biased. Curiously, however, K&K seem to imply that because true detection function may not necessarily satisfy the RSPF condition, single-visit methodology should not be used in practice. Our disagreement is with this implication.

Such an implication is very strange on two accounts. First, it is well known that validity of *every* statistical method depends on assumptions. The results are sensitive to the violation of these assumptions. Second, and more importantly, whether the *true* generating mechanism satisfies all the assumptions can seldom be known. For example, ecologists use maximum likelihood estimators (MLE) and the associated confidence intervals that are based on the result that the MLEs are consistent and asymptotically normal. This result holds true only if a number of regularity conditions are satisfied by the underlying *true* mechanism. Aside from the basic requirement that the model parameters are identifiable, the regularity conditions relate to the expected value of the higher order derivatives of the log-likelihood function. In the dependent data situation, one needs that the underlying true mechanism is a $\phi$-mixing process of certain order. Population time series analysis is a common research activity in quantitative ecology. We are not aware of any papers that prove that the *true* underlying mechanism satisfies this mixing condition and other regularity conditions. Hierarchical models are also commonly used to analyze ecological data. These models make distributional assumptions about the latent variables. Statistical inferences are sensitive to the distributional assumptions on latent variable and in most cases one cannot test the validity of the latent variable model specification. In many cases, these latent variables do not even correspond to any observable characteristics and hence there is not even a potential to test the assumptions in practice. We can continue with this list *ad infinitum* but the moral of the story is that it is not a news that statistical methods are sensitive to model misspecification. Moreover, one cannot always test the validity of the model specification.

We do not take an issue with K&K's mathematical findings that single-visit method is sensitive to the RSPF condition. As we showed above, one can launch such a criticism against every statistical method that has ever been proposed. We note that the multiple-visit N-mixture (MV) method is not criticized for simply being sensitive to the closed population assumption; it is criticized because the closed population assumption is seldom satisfied *in practice* (Bayne et al. 2011, Chandler et al. 2011, Dail and Madsen 2011). If this assumption were satisfied in most situations, we would be using MV method without qualms. It is well known that the standard bootstrap method fails if, among many other regularity conditions, the underlying *true* distribution is heavy tailed (Athreya, 1987). In spite of the possibility that the underlying true distribution may be heavy tailed, we continue to use the bootstrap method *in practice* because our experience suggests that such situations are *rare*.

Vast scientific experience, not only in the field of detection error but also in various other scientific ventures, of modeling probability of an event suggests that most probability models that are actually used in practice do satisfy the RSPF condition. Hence we claim that the detection error models are more likely to satisfy the RSPF condition than not and that we are likely to be correct with our inferences.

As we have already acknowledged, single-visit methods are sensitive to the violation of the RSPF condition. There are two ways to deal with the issue of sensitivity. One is to be aware of it and accept the (hopefully, small) probability that the results are potentially wrong. The other is to modify the method to make it robust against such deviations. Population closure is considered unlikely in practice and hence methods are being developed that are robust against this assumption. Single-visit (SV) method replaces the population closure assumption with the RSPF condition. Generalized (or dynamic) N-mixture method (DM; Dail & Madsen 2011) replaces the population closure assumption by explicitly modeling the population changes from one time point to other. Both these assumptions are liable to fail. For example, underlying population changes occur on continuous time scale and there are many different mechanisms. DM models the population change on discrete time scale and with a particular form of transition matrix. There is no way to check the appropriateness of this discrete time population transition model given the observed data. If it is not appropriate, DM will lead to biased estimates. It is simply unrealistic to make statistical methods robust against every possible model misspecification. It is an occupational hazard that any time a statistical method is used to analyze real data, there is always a (hopefully small) possibility that the true data generating mechanism is such that assumptions are not satisfied and hence the results are completely wrong.

*Sensitivity of the generalized N-mixture method to scaled probability link*

K&K show that when detection probability is modeled as scaled logistic, the SV leads to biased estimators. K&K, citing Zipkin et al. (2014), suggest that the generalized N-mixture model might have numerical issues under the standard models. However, the behaviour of the model was not studied using scaled link functions. To fill this gap, we present a simulation study to better understand the behaviour of DM under various situations. We conducted the simulation study so that the stated assumptions of the DM are fully satisfied. We follow the models and notation described in Dail and Madsen (2011). We used two continuous covariates that affected abundance and two continuous covariates that affected detectability. In one setup there was no overlap between the abundance and detection covariates. In the other setup, the covariates were overlapping: one of the continuous covariates affected both abundance and detection. For the transition parameters, we used constant 'arrival' $\gamma = 1$ and constant 'survival' $\omega$ values ($\omega$ ranged from 0 to 1 in steps of 0.1). The assumption that transition parameters do not depend on covariates is not realistic. It implies that as time passes, the abundances are less affected by the covariates. However, when these parameters do depend on covariates, the estimation procedure is extremely slow and extensive simulations are nearly impossible to perform. Because the dependence on the covariates vanishes for other time points, we compared the estimated mean abundances under DM and SV only for the first visit where the abundances do depend on covariates. We considered 200 locations and 3 replicate visits. This is a fairly large sample for carrying out multiple visits in practice. For smaller number of locations but more replicate visits, the biases are even larger. We used the 'unmarked' package (Fiske & Chandler 2011) to estimate the parameters for DM and the 'detect' package (Sólymos, 2012) to analyse the single-visit data. Reproducible R (R Core Team 2014) simulation code can be found in the Supporting Information.

Let us look at the results (Figure 3) when the RSPF condition is satisfied, that is, when $1/c = 1$ ($c \geq 1$ is the scaling factor in the logit link following the notation of K&K). In this case, both DM and SV methods are nearly unbiased. DM method exhibits increasing negative bias as the populations become more open ($\omega < 1$) and when scaled logistic model ($1/c < 1$) is used for detection probability. In fact, if $w = 0$, DM likelihood is identical to the SV likelihood and hence, it also requires that the detection function satisfies the RSPF condition. This is reflected in the fact that the pattern in the DM based bias is identical to the pattern in single-visit based bias when the dependence between consecutive visits is weak.

An obvious conclusion from this simulation study is that DM is also not robust against the scaled logistic detection function. Of course, no simulation study can ever determine that one method is better than other methods under every possible scenario. We have always presented, and still consider, single-visit method as one of the tools in the statistical toolbox available to the ecologists. It is certainly an improvement over the multiple-visits N-mixture method and is definitely a reasonable alternative to the generalized N-mixture method when the dynamic parameters are unknown or when time-series data are unavailable.

*Diagnosing the presence of scaled probability links*

For a scientist and a statistician, a mathematical *model* is not simply a mathematical formula. A mathematical or a statistical model should represent a realistic mechanism. Neither K&K nor Hastie and Fithian (2013) offer a mechanism that might underlie a scaled probability link function. In this section, we propose a few possible mechanisms that could lead to scaled probability models. Furthermore, we show how single visit datasets can be used to diagnose the possibility of scaling in the detection function. A constant scaling model by itself is not very useful in practice because it does not further our understanding of the detection process. In order for a mathematical model to be useful, we should be able to use it not only for understanding the process of detection but also to use that knowledge in designing effective surveys. This suggests that we should try to find out if there are any covariates that affect the scaling factor. Hence we show how such covariates can be incorporated in the scaling component of the detection model. It again turns out that the RSPF condition becomes important for identifiability of covariate dependent scaling factors.

A constant scaling factor in the detection function or a resource selection probability function may arise because of data entry errors. For example, occasionally an observer, being human, might transcribe 'absent' even when he means to write 'present'. Proportion of such random data entry errors (to be precise, 1 – proportion of errors) are represented as a constant scaling factor in the detection function. Although we hope such errors are infrequent in practice. Similarly, in telemetry studies, occasionally one may miss a GPS location completely at random; that is, the probability of missingness does not depend on the habitat the animal is present (Frair et al. 2004). Again, such errors are rare in practice but can be represented as a constant scaling factor in the RSPF model. It is much more likely that the probability of missingness depends on covariates. We will show how to deal with that case later in the section.

We first start with an approach to diagnose the presence of a scaling factor in the detection function when ancillary information at a subset of locations is collected. We emphasize that this information can be collected during the single visit surveys without any need to revisit the location repeatedly. Thus logistical requirements are similar to the single visit surveys.

*Distance sampling extension of the Binomial-ZIP model*

It often happens that a subset of the data is collected using a different protocol. For example, when professional biologists are conducting surveys at a few locations as part of an otherwise volunteer based program, they may collect information about distance classes within which the individuals are observed. It is known that the inferences are highly sensitive to the correct estimation of the distances. Only a highly trained field staff can obtain reliable information on the distances at a subset of the locations in the larger survey.

For this subset of the locations, the total count $Y_i$ at site $i$ is the sum of the vector $(Y_{i1}, \ldots, Y_{iJ})$ with individuals detected in different distance bands or strips ($j = 1, \ldots, J$). We can write the following hierarchical model to allow for this stratification by distances:

$$N_i \sim \text{Poisson}(\lambda_i)$$

$$(O_i | N_i) \sim \text{Binomial}(N_i, p_i)$$

$$(Y_{i0}, Y_{i1}, \ldots, Y_{iJ} | O_i) \sim \text{Multinomial}(O_i, (\pi_{i0}, \pi_{i1}, \ldots, \pi_{iJ}))$$

Where $(\pi_{i0}, \pi_{i1}, \ldots, \pi_{iJ})$ is the unconditional Multinomial cell probability vector, which can be a function of distance (see Sólymos et al. 2013). The count $Y_{i0}$ is not observed with a corresponding cell probability $\pi_{i0} = 1 - \sum_{j=1}^{J} \pi_{ij} = 1 - q_i$. A constant truncation distance across the survey locations leads to constant $q_i = q$ for all locations. This Multinomial model, thus, corresponds to the constant scaling factor model proposed by K&K. Moreover, the marginal distribution of $Y_i$, the total counts at the site, is a Binomial-Poisson mixture with detection probability $p_i q$, a scaled probability function. Thus, any constant scaling of $p_i$ gets absorbed in $q$. In practice, to account for large number of zeros, it is often sensible to model $N_i$ as a zero-inflated Poisson random variable. As in the single-visit data (Sólymos et al. 2012), it is known that the zero-inflation parameter and the detection error parameters tend to be confounded. In Appendix 1, we extend the conditional likelihood method to the Multinomial-ZIP model. This analysis can be conducted using the function 'svabuRDm' in the Supporting Information.

Analysis of the data using only the marginal distribution of $Y_i$ will lead to biased estimates of the intercept parameter for the abundance model. On the other hand, analysis based on the Multinomial model will lead to unbiased estimates of the same. Hence the analysis of such a subset of the data can be used to check the assumption that the detection function satisfies the RSPF condition. One can predict the abundances at this subset of locations using two methods: SV method and the Multinomial method. If the predicted abundances are substantially different, scaled probability function might be needed for the detection.

To demonstrate that our proposed Multinomial model can be used to estimate abundance in the presence of an unknown scaling factor, we performed a simulation study to compare the relative bias in abundance intercept between the 3-level Multinomial model and the single-visit method. We used the scaled logistic detection probability function with scaling factor $q = 0.5$ corresponding to 100 m truncation distance and $\tau = 80$ m effective detection radius. Results in Figure 4 indicate that the 3-level Multinomial model gives nearly unbiased abundance intercept values. We know that the single-visit method would lead to biased estimates of the abundance. The SV abundance intercept parameter shows a bias approximately equal to $\log(q) = \log(0.5) = -0.69$ (Figure 4). Comparison of results between the 3-level model and the single-visit model would alert the researcher about the need to consider a scaled logistic model. When extra information on the distance classes is available for a subset of locations, it is possible to exploit such extra information to test for scaling and use the estimates as offsets in single-visit models (see Sólymos et al. 2013 for a discussion of detectability offsets).

*Incorporating covariates in the scaling function*

Suppose a researcher realizes the need for incorporating a scaling factor in the detection function using the diagnostic tool described above or based on their understanding of the detection process. In our opinion, a detection function is not a nuisance parameter that only needs to be estimated to correct the abundance estimates. It should also be used to help design effective surveys in the future. A constant scaling factor is not of much use for such a purpose. One needs to understand why there is a scaling factor and how one can control its value and potentially increase the detection probability. Given the results in Welsh et al. (2013), the goal should be to design surveys so that detection probability is high and correcting the abundance estimates is less important. Hence, after diagnosing the presence of the scaling factor, one should strive to model it as a function of covariates. We give an example of such a situation and show how to fit the scaled detection function that depends on covariates. Not surprisingly, the RSPF condition again becomes important for identifiability. Either the scaling function or the original detection function needs to satisfy the RSPF condition.

In distance sampling, it is often assumed that the detection function takes value 1 at distance 0 (Buckland et al. 2001, Matsuoka et al. 2012). This has been criticized as being not always realistic (Johnson 2008) and one can imagine a situation where an upper limit to the detection probability that is smaller than 1 exists (Farnsworth et al. 2002, Sólymos et al. 2013, Amundson et al. 2014). It is, however, difficult to imagine that the upper limit for detection probability is constant under all situations. It is far more likely that the upper limit depends on habitat covariates or some measure of sampling effort (Reidy et al. 2011, Sólymos et al. 2013, Matsuoka et al. 2014, Kellner & Swihart 2014). As we show below, under this more realistic situation, the single-visit method leads to identifiable parameters and correctly provides abundance estimates. Consider a situation where we are surveying forest songbirds. Suppose there are $N$ birds at a location. We can model these by the Poisson distribution (or, Negative Binomial or zero-inflated versions of these). Not all of them might be singing during the survey time interval. The birds that are singing are available for detection. This may be modeled by using a Binomial distribution that is related to the proportion of birds singing. Actual detection, however, is still imperfect because even if the bird is singing, the observer may not detect it with probability 1 depending, for example, on the distance between the observer and the vocalizing bird. This leads to another hierarchy that can be modeled by a Binomial distribution that is related to detection. The full model may be written as:

$$N_i \sim \text{Poisson}(\lambda_i)$$

$$(O_i|N_i) \sim \text{Binomial}(N_i, p_i)$$

$$(Y_i|O_i) \sim \text{Binomial}(O_i, q_i)$$

Notice that the distribution of the observed data given the true counts reduces to $(Y_i|N_i) \sim \text{Binomial}(N_i, p_i q_i)$. If we take $q_i = q$ to be independent of any covariates, this reduces to the scaled detection probability mentioned by K&K. However, it is far more likely that the scaling factor depends on covariates. The method of conditional likelihood (Sólymos et al. 2012) can be extended to this case. Mathematical details are available in the Appendix 2. Analysis of such data can be conducted using the function 'svabuRD' in the Supporting Information. In Figure 5, we present simulation results showing that when $q_i$ depends on a covariate, such as point count radius, all the parameters are estimable using the single-survey.

**Conclusions**

All statistical analyses inherently depend on models and assumptions. As practicing scientists, we are well aware of the reality that we never know whether or not the true data generating mechanism satisfies these assumptions. In spite of this, we continue to use statistical methods for data analysis as long as the assumptions are not unrealistic. We consider the risk of being wrong occasionally as simply an occupational hazard.

In this paper, we have clarified the conditions under which single-survey methodology is valid. We argued that the situations under which the RSPF and single-visit methodologies fail are rare. This is further emphasized by the lack of any references to published studies where the log or scaled logistic model is preferred over the commonly used link functions for the probability of detection. We also illustrated that the generalized N-mixture method does not protect against the violation of the RSPF condition. If the population is completely open ($\omega = 0$), DM likelihood reduces to the single-visit N-mixture likelihood, and as such, parameters are non-identifiable under the DM method unless the detection function satisfies the RSPF condition. We discuss various mechanisms that may lead to scaling in the detection function. We discuss a practical method, based only on a single visit to the survey locations, to diagnose the presence of a scaling factor. If the scaling factor is substantially different than 1, one may need to think about possible mechanisms for scaling. We showed that if the scaling factor is a function of covariates, single-visit methodology is useful to estimate the parameters in the scaling function.

In practice, when designing surveys, one has to strike a balance between the increase in cost due to multiple-visits, possibility of failure of critical assumptions, such as independence of surveys and the possibility that the RSPF condition is not satisfied by the true model of detection when aiming to make statements about abundance and occupancy. Even a cursory look at the current literature suggests that the detection models that are actually used do belong to the RSPF model class. This fact and the breadth of the RSPF model class suggest that one may be on safe grounds to use single visit method for data analysis and making inferences.

**Acknowledgements**

SRL was partially supported by the NSERC Discovery grant. Comments from E. Bayne, J. Wright, the Associate Editor, an anonymous reviewer, and J. Knape on earlier versions improved the paper. This research was enabled in part by support provided by WestGrid (www.westgrid.ca) and Compute Canada Calcul Canada (www.computecanada.ca).

**Supporting information**

The code to reproduce the simulations and results presented in this paper are provided at: https://github.com/psolymos/detect/tree/master/extras/revisitingSV.

**References**

Amundson, C. L., Royle, J. A., & Handel, C. M. (2014) A hierarchical model combining distance sampling and time removal to estimate detection probability during avian point counts. *The Auk: Ornithological Advances*, 131, 476–494.

Athreya, K. B. (1987) Bootstrap of the mean in the infinite variance case. *Annals of Statistics*, 15, 724–731.

Bayne, E. M., Lele, S. R., & Sólymos, P. (2011) *Bias in estimation of bird density and relative abundance when the closure assumption of multiple survey approaches is violated: a simulation study*. Technical report, Boreal Avian Modelling Project. pp. 26. URL: http://www.borealbirds.ca/files/Bayne_et_al_2011_Bias_In_Estimation_Rpt.pdf [accessed January 18 2015]

Buckland, S.T., Anderson, D.R., Burnham, K.P., Laake, J.L., Borchers, D.L. & Thomas, L. (2001) *Introduction to Distance Sampling: Estimating Abundance of Biological Populations*. OxfordUniversity Press,Oxford,UK.

Campbell, D., & Lele, S. R. (2014) An ANOVA test for parameter estimability using data cloning with application to statistical inference for dynamic systems. *Computational Statistics & Data Analysis*, 70, 257-267.

Chandler, R. B., Royle, J. A., & King, D. I. (2011) Inference about density and temporary emigration in unmarked populations. *Ecology*, 92, 1429–1435.

Cook, R. J. & Lawless, J. (2007) *The statistical analysis of recurrent events*. Springer, Series: Statistics for Biology and Health.

Dail, D. & Madsen, L. (2011) Models for estimating abundance from repeated counts of an open metapopulation. *Biometrics*, 67, 577–587.

Frair, J. L., Nielsen, S. E., Merrill, E. H., Lele, S. R., Boyce, M. S., Munro, R. H. M., Stenhouse, G. B., & Beyer, H. L. (2004) Removing GPS collar bias in habitat selection studies. *Journal of Applied Ecology*, 41, 201–212.

Farnsworth, G.L., Pollock, K.H., Nichols, J.D., Simons, T.R., Hines, J.E., & Sauer, J.R. (2002) A removal model for estimating detection probabilities from point count surveys. *Auk*, 119, 414–425.

Fiske, I., & Chandler, R. (2011) unmarked: an R package for fitting hierarchical models of wildlife occurrence and abundance. *Journal of Statistical Software*, 43, 1–23. URL: http://www.jstatsoft.org/v43/i10/.

Gilbert, P., Lele, S. R., & Vardi, Y. (1999) Maximum likelihood estimation in semiparametric selection bias models with application to AIDS vaccine trials. *Biometrika*, 86, 27–43.

Hastie, T. & Fithian, W. (2013) Inference from presence-only data; the ongoing controversy. *Ecography*, 36, 864–867.

Johnson, D.H. (2008) In defense of indices: the case of bird surveys. *Journal of Wildlife Management*, 72, 857–868.

Kellner, K. F., & Swihart, R. K. (2014) Accounting for imperfect detection in ecology: A quantitative review. *PLoS ONE*, 9, e111436.

Knape, J., & Korner-Nievergelt, F. (2015) Estimates from non-replicated population surveys rely on critical assumptions. *Methods in Ecology and Evolution*, in press.

Lele, S.R., Nadeem, K., & Schmuland, B. (2010) Estimability and likelihood inference for generalized linear mixed models using data cloning. *Journal of the American Statistical Association*, 105, 1617–1625.

Lele, S.R., & Allen, K. L. (2006) On using expert opinion in ecological analyses: a frequentist approach. *Environmetrics*, 17, 683–704.

Lele, S. R., & Keim, J. L. (2006) Weighted distributions and estimation of resource selection probability functions. *Ecology*, 87, 3021–3028.

Lele, S.R., Moreno, M., & Bayne, E. (2012) Dealing with detection error in site occupancy surveys: What can we do with a single survey? *Journal of Plant Ecology*, 5, 22-31.

Lele, S. R., Merrill, E. H., Keim, J., & Boyce, M. S. (2013) Selection, use, choice and occupancy: clarifying concepts in resource selection studies. *Journal of Animal Ecology*, 82, 1183-1191.

MacKenzie, D.I., Nichols, J.D., Lachman, G.B., Droege, S., Royle, J.A., Langtimm, C.A. (2002) Estimating site occupancy rates when detection probabilities are less than one. *Ecology*, 83, 2248–2255.

Matsuoka, S.M., Bayne, E.M., Sólymos, P., Fontaine, P., Cumming, S.G., Schmiegelow, F.K.A., & Song, S.J. (2012) Using binomial distance-sampling models to estimate the effective detection radius of point-count surveys across boreal Canada. *Auk*, 129, 268–282.

Matsuoka, S. M., Mahon, C. L., Handel, C. M., Sólymos, P., Bayne, E. M., Fontaine, P. C., & Ralph, C. J. (2014) Reviving common standards in point-count surveys for broad inference across studies. *Condor*, 116, 599–608.

Moreno, M., & Lele, S. R. (2010) Improved estimation of site occupancy using penalized likelihood. *Ecology*, 91, 341-346.

R Core Team (2014) *R: A language and environment for statistical computing*. R Foundation for Statistical Computing, Vienna, Austria. URL: http://www.R-project.org/.

Reidy, J. L., Thompson III, F. R., & Bailey, J. W. (2011) Comparison of methods for estimating density of forest songbirds from point counts. *Journal of Wildlife Management*, 75, 558–568.

Rota, C.T., Fletcher, R.J., Dorazio, R.M., & Betts, M.G. (2009) Occupancy estimation and the closure assumption. *Journal of Applied Ecology*, 46, 1173–1181.

Royle, J.A. (2004) N-mixture models for estimating population size from spatially replicated counts. *Biometrics*, 60, 108–115.

Royle, J. A., Dawson, D. K., & Bates, S. (2004) Modeling abundance effects in distance sampling. *Ecology*, 85, 1591–1597.

Sólymos, P. (2010) dclone: data cloning in R. *R Journal*, 2, 29–37.

Sólymos, P., Lele, S.R., & Bayne, E. (2012) Conditional likelihood approach for analyzing single visit abundance survey data in the presence of zero inflation and detection error. *Environmetrics*, 23, 197–205.

Sólymos, P., Moreno, M., & Lele, S.R. (2014) detect: analyzing wildlife data with detection error. R package version 0.3-2. http://dcr.r-forge.r-project.org/.

Sólymos, P., Matsuoka, S. M., Bayne, E. M., Lele, S. R., Fontaine, P., Cumming, S. G., Stralberg, D., Schmiegelow, F. K. A., & Song, S. J. (2013) Calibrating indices of avian density from non-standardized survey data: Making the most of a messy situation. *Methods in Ecology and Evolution*, 4, 1047–1058.

Welsh, A.H., Lindenmayer, D. B., & Donnelly, C. F. (2013) Fitting and interpreting occupancy models. *PLoS ONE*, 8, e52015.

Zipkin, E. F., Thorson, J. T., See, K., Lynch, H. J., Grant, E. H. C., Kanno, Y., Chandler, R. B., Letcher, B. H. & Royle, J. A. (2014) Modeling structured population dynamics using data from unmarked individuals. *Ecology*, 95, 22–29.

**Figure legends**

**Figure 1**. The variety of randomly generated linear, quadratic, and cubic response curves after inverse logit and cloglog transformations illustrate that the class of models that satisfy the RSPF condition (as described in the text) is fairly general.

**Figure 2**. Complex nonlinear response functions that do not reach 1 are identifiable as long as the RSPF condition explained in the text is satisfied. Grey lines show results for best fit response curves from 100 simulations based on Resource Selection Probability Function (RSPF) model (left) and single-visit N-mixture model (right) using one of the cubic polynomial response functions (black curve) from Figure 1. Best fit curve was selected based on AIC values from linear, quadratic and cubic responses; cubic response was supported in 100% of the simulation runs. Sample sizes were: 3000 used and 30000 available points in RSPF and 1000 observations in N-mixture.

**Figure 3**. Relative bias [(estimate – true value) / true value] in mean abundance ($\bar{\lambda}$) estimates based on generalized N-mixture (DM) and single-visit N-mixture (SV) models as a function of the scaling constant ($1/c$) used in the scaled logistic link. Blue-to-red colored lines indicate different values of the survival rate ($\omega$), models exhibited trend due to a fixed arrival rate ($\gamma = 1$) through the $T = 3$ repeated visits (only estimates for the first visit shown, see Supporting Information for R code and full set of results). The set of abundance and detection covariates were disjunct (top row) or included a variable in common (bottom row). Lines represent mean values from 120 replicates. Number of locations was 200.

**Figure 4**. Bias (estimate – true value) in abundance intercept sestimates based on 3-level Multinomial (Multinom) and single-visit N-mixture (SV). A scaling constant of $q = 0.5$ was used in the scaled logistic link for simulations. Results are based on 100 simulation runs, $n = 200$ survey locations, and two distance bands (0–50 m and 50–100 m bands around points).

**Figure 5**. Estimability check for the single-visit Binomial–Binomial–Poisson mixture to estimate parameters of three processes: population density ($\alpha_0$, $\alpha_1$), singing behaviour of birds (availability; $\beta_0$, $\beta_1$), and distance related observation process (detectability; effective detection radius, $\tau$). Left panel shows bias for the five model parameters. The middle shows the detection function related to the $\tau$ parameter (black line is the true relationship, grey lines represent replicate runs). The bias for mean abundance ($\bar{\lambda}$), mean availability ($\bar{p}$), and mean probability of detection ($\bar{q}$) corresponding to the three processes are shown in the right panel. Box-plots represent simulation results based on 100 replicates. Bias was calculated as (estimate – true value), relative bias was defined as [(estimate – true value) / true value].

**Appendix 1**

**Conditional maximum likelihood estimation of the Multinomial–ZIP model parameters**

Let us consider an extension of the single-visit abundance methodology developed in Sólymos et al. (2012) to the case where the scaled link function will be useful. Imagine a survey situation where the observer is counting birds within a specified radius circle at each survey location. The bird has to make an auditory or visual cue in order to be detected. This is often called availability for sampling (Reidy et al. 2011). One can model the availability by using the standard links such as logit or cloglog as a function of covariates such as time of the day, time of year, wind, precipitation etc. In the following, we denote this by $p_i$, probability of availability at the $i$-th location. However, even if an individual is available for observation, it does not mean it will be observed. This probability may depend on how far the bird is from the observer. A common approach in distance sampling is to bin the detections according to distance classes. This data corresponds to a Multinomial distribution with counts $(Y_{i1}, \ldots, Y_{iJ})$ observed in different distance strata ($j = 1, \ldots, J$) with corresponding cell probabilities $(p_i \pi_{i1}, \ldots, p_i \pi_{iJ})$. Availability, $p_i$, can be constant or a function of covariates, while $\pi_{ij}$ can be a function of distance (see Royle et al. 2004 and Sólymos et al. 2013). It is required to have at least 2 distance intervals (e.g. 0–50 m and 50–100 m) with *finite* truncation distance. Because we have more information in the data when a single-visit count is subdivided by spatial stratification, the truncation distances ($r_{iJ}$) do not need to vary across locations. We show that the conditional distribution of the observations given the total count is greater than zero does not depend on the zero inflation parameter ($\varphi$) of the ZIP abundance model. The conditional distribution may be written as:

$$P(Y_{i\cdot} = y_{i\cdot} | Y_{i\cdot} > 0) = \frac{P(Y_{i\cdot} = y_{i\cdot})}{1 - P(Y_{i\cdot} = 0)}$$

where $y_{i\cdot} = \sum_{j=1}^{J} y_{ij}$. The probability mass function for this probability distribution is given by:

$$P(Y_{i\cdot} = y_{i\cdot} | Y_{i\cdot} > 0) = \frac{\sum_{N_i=y_{i\cdot}}^{\infty} \frac{N_i!}{y_{i0}! y_{i1}! \dots y_{iJ}!} \pi_{i0}^{y_{i0}} \{p_i \pi_{i1}\}^{y_{i1}} \dots \{p_i \pi_{iJ}\}^{y_{iJ}} e^{-\lambda_i} \frac{\lambda_i^{N_i}}{N_i!}}{1 - e^{-\lambda_i p_i q_i}}$$

where $y_{i0} = N_i - \sum_{j=1}^{J} y_{ij} = N_i - y_{i\cdot}$ is the unobserved count portion of the unknown variable $N_i$ ($y_{i0} = N_i$ when $y_{i\cdot} = 0$); and $\pi_{i0} = 1 - \sum_{j=1}^{J} p_i \pi_{ij} = 1 - p_i q_i$ is the corresponding cell probability. The derivation follows along the same lines as in Solymos et al. (2012).

$$P(Y_{i\cdot} = y_{i\cdot}) = (1 - \varphi) \sum_{N_i=y_{i\cdot}}^{\infty} \frac{N_i!}{y_{i0}! y_{i1}! \dots y_{ik}!} \pi_{i0}^{y_{i0}} \{p_i \pi_{i1}\}^{y_{i1}} \dots \{p_i \pi_{iJ}\}^{y_{iJ}} e^{-\lambda_i} \frac{\lambda_i^{N_i}}{N_i!}$$

$$P(Y_{i\cdot} = 0) = \varphi + (1 - \varphi) \sum_{N_i=0}^{\infty} \frac{N_i!}{y_{i0}! 0! \dots 0!} \pi_{i0}^{y_{i0}} \{p_i \pi_{i1}\}^{0} \dots \{p_i \pi_{iJ}\}^{0} e^{-\lambda_i} \frac{\lambda_i^{N_i}}{N_i!}$$

$$= \varphi + (1 - \varphi) \sum_{N_i=0}^{\infty} \frac{N_i!}{N_i!} \pi_{i0}^{N_i} 1 \dots 1 e^{-\lambda_i} \frac{\lambda_i^{N_i}}{N_i!}$$

$$= \varphi + (1 - \varphi) \sum_{N_i=0}^{\infty} \pi_{i0}^{N_i} e^{-\lambda_i} \frac{\lambda_i^{N_i}}{N_i!}$$

$$= \varphi + (1 - \varphi) e^{-\lambda_i} \sum_{N_i=0}^{\infty} \pi_{i0}^{N_i} \frac{\lambda_i^{N_i}}{N_i!}$$

$$= \varphi + (1 - \varphi) e^{-\lambda_i} \sum_{N_i=0}^{\infty} (1 - p_i q_i)^{N_i} \frac{\lambda_i^{N_i}}{N_i!}$$

$$= \varphi + (1 - \varphi) e^{-\lambda_i} \sum_{N_i=0}^{\infty} \frac{[(1 - p_i q_i) \lambda_i]^{N_i}}{N_i!}$$

$$= \varphi + (1 - \varphi) e^{-\lambda_i} e^{(1 - p_i q_i) \lambda_i}$$

$$= \varphi + (1 - \varphi) e^{-\lambda_i p_i q_i}$$

Hence, $1 - P(Y_{i\cdot} = 0) = (1 - \varphi)(1 - e^{-\lambda_i p_i q_i})$. The Supporting Information provides code to simulate data under this model and estimate the parameters.

**Appendix 2**

**An example where scaled link function for detection probability is applicable**

Here we use a distance sampling approach similar to Appendix 1 and consider a half-Normal detectability function to model the effect of the distance. Following Sólymos et al. (2013), this probability may be modelled as the function of the point count radius ($r_i$) and the standard deviation of the half-Normal distribution ($\tau$) (see equation on p. 1050 in Sólymos et al. 2013). The point count radius corresponds to the survey protocol that decides the circle within which the observer counts the birds. This may vary from station to station or from survey to survey. However, it is important that it varies. We can write the marginal probability mass function as

$$P(Y_i = y_i) = \sum_{N_i=y_i}^{\infty} \binom{N_i}{y_i} (p_i q_i)^{y_i} (1 - p_i q_i)^{N_i - y_i} e^{-\lambda_i} \frac{\lambda_i^{N_i}}{N_i!}$$

where $\lambda_i = D_i \pi r_i^2$ and $D_i$ is the population density that we model using log-link and covariates. The marginal distributions under the Zero-inflated Poisson model and corresponding Negative Binomial models can be derived in a similar fashion. It is also trivial to extend the method of conditional likelihood (Sólymos et al. 2012) to this case using the fact that conditional distribution $P(Y_i = y_i | Y_i > 0)$ is independent of the zero inflation parameter for when the distribution of $N_i$ is zero inflated. For example, for zero inflated Poisson model it is given by

$$P(Y_i = y_i | Y_i > 0) = \frac{\sum_{N_i=y_i}^{\infty} \binom{N_i}{y_i} (p_i q_i)^{y_i} (1 - p_i q_i)^{N_i - y_i} e^{-\lambda_i} \frac{\lambda_i^{N_i}}{N_i!}}{1 - e^{-\lambda_i p_i q_i}}$$

Under the RSPF condition and provided that covariates that affect '$p$' and '$q$' are not completely overlapping, this model leads to identifiable parameters. See Lele et al. 2012 and the present paper for the discussion of identifiability for single-visit Binomial-Binomial mixture, which is exactly the same as the $p_iq_i$ component discussed here. The Supporting Information provides code to simulate data under this model and estimate the parameters.

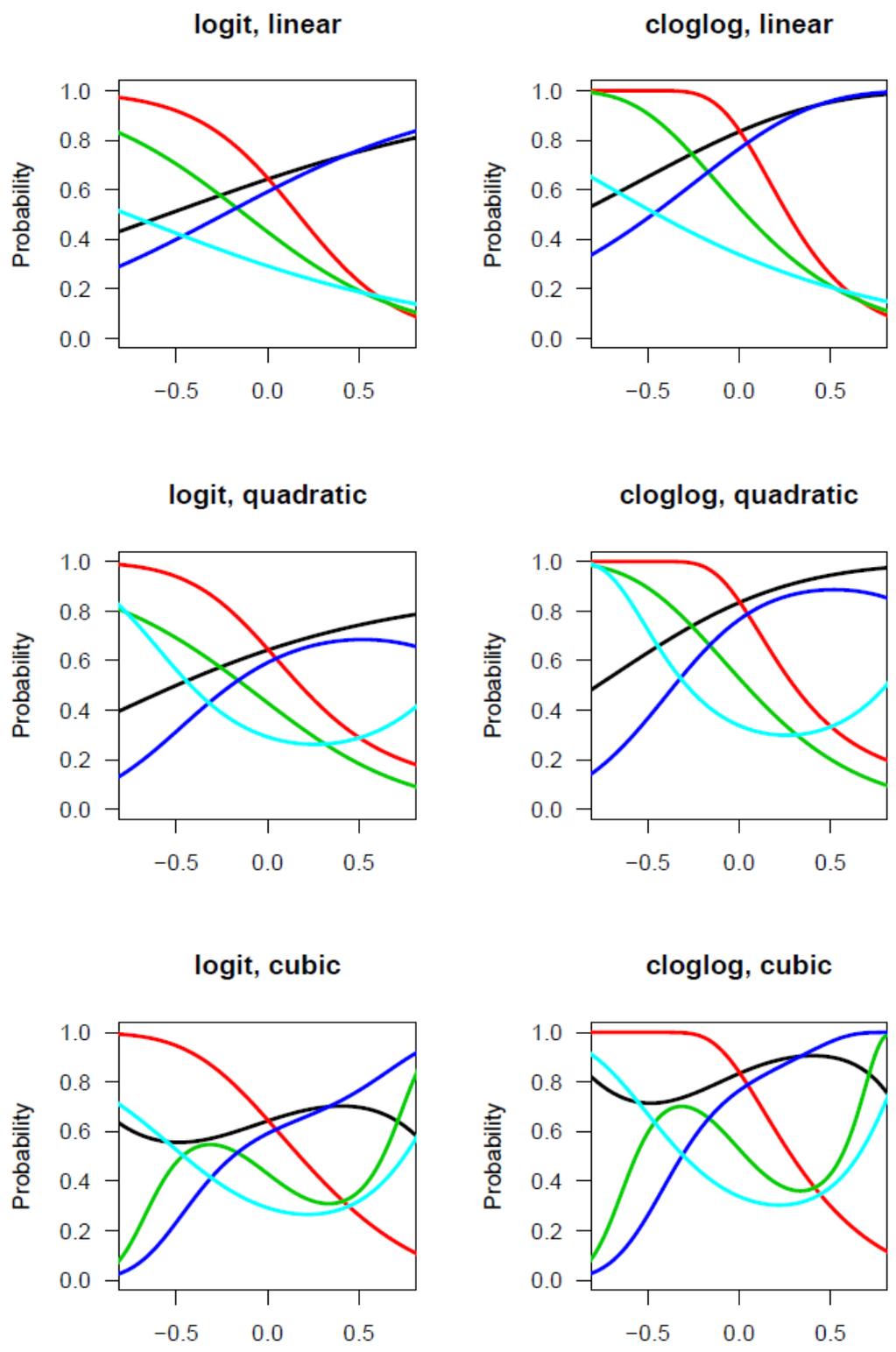

Fig. 1

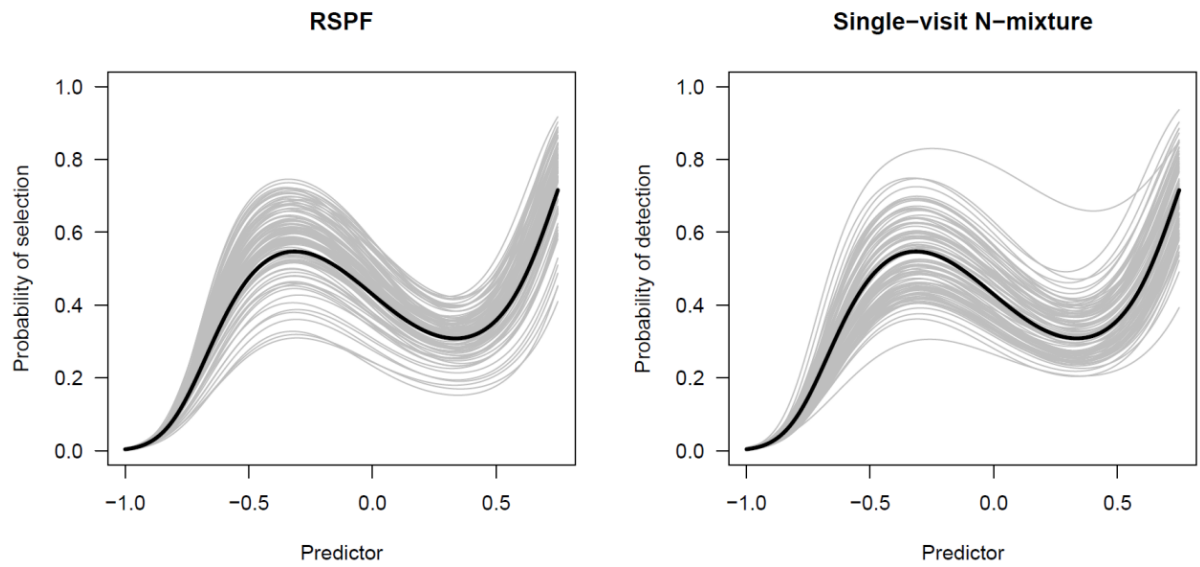

Fig. 2

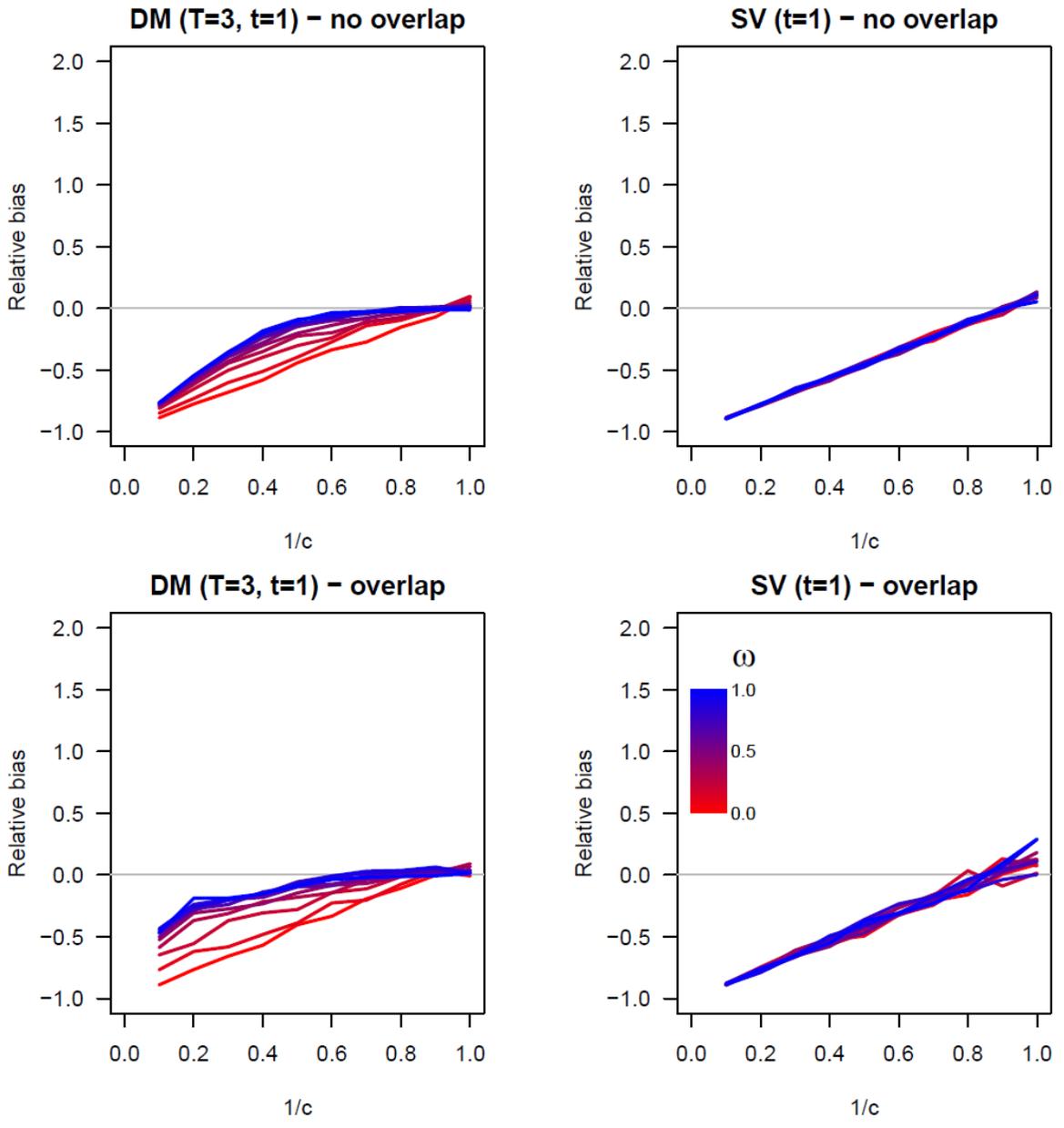

Fig. 3

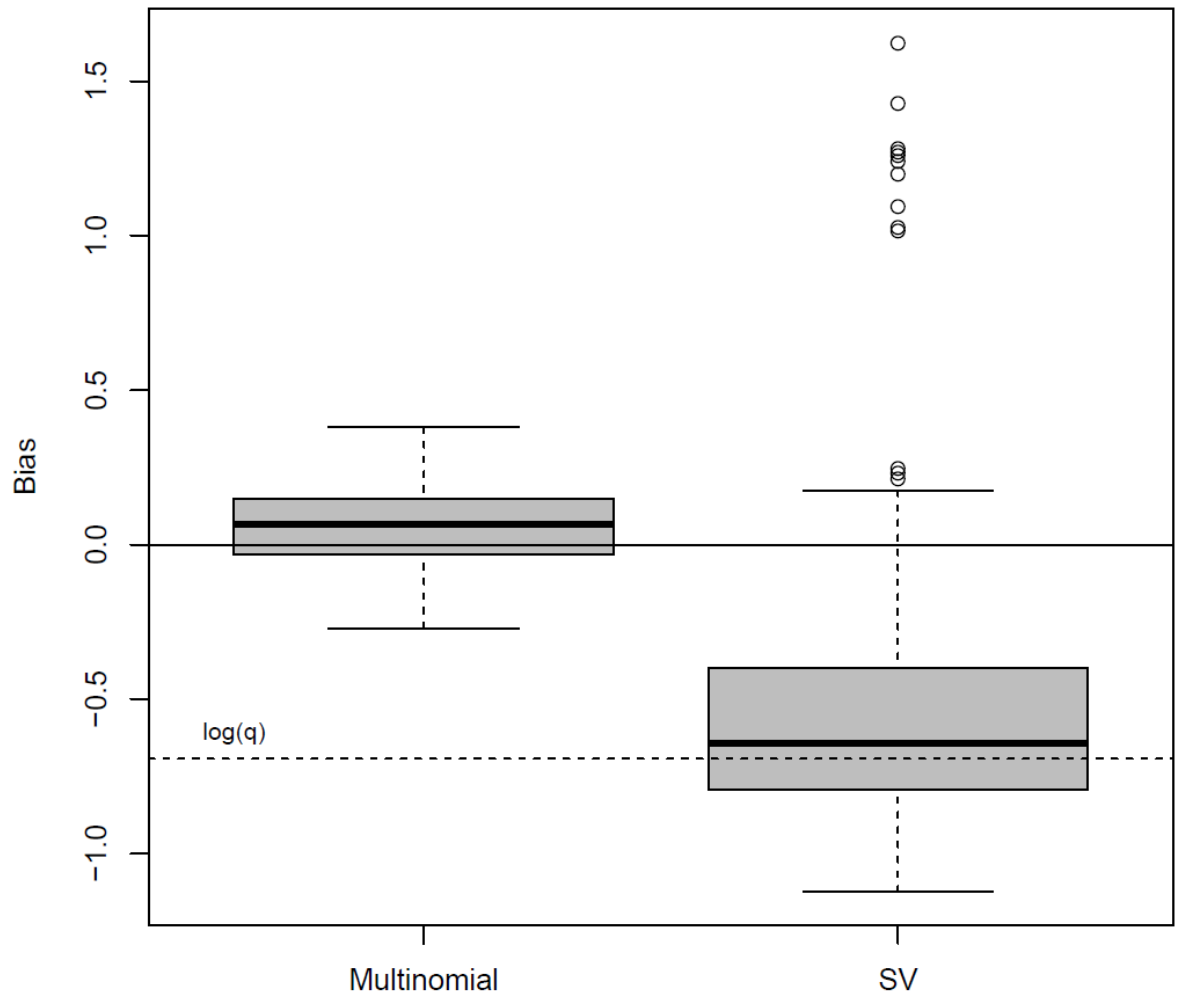

Fig. 4

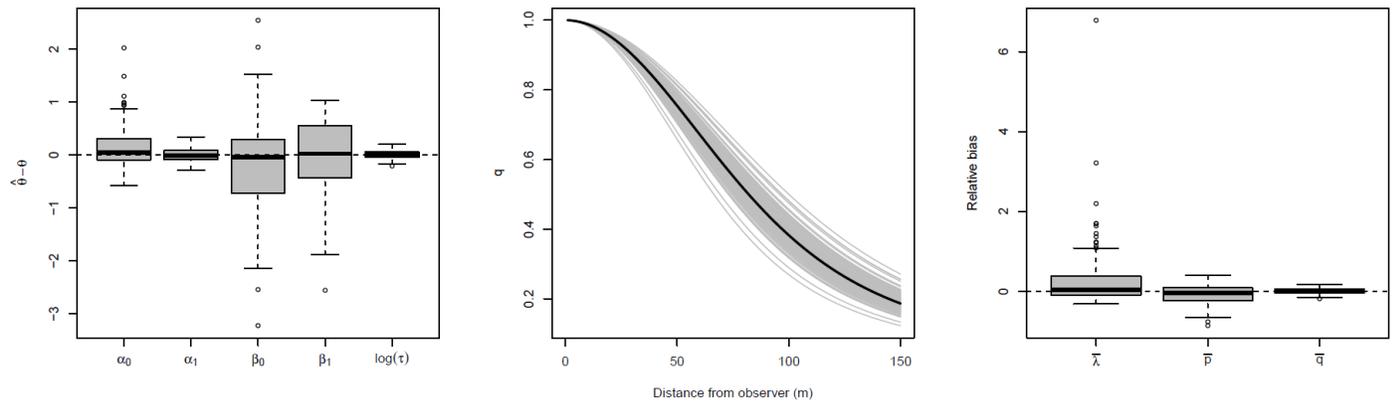

Fig. 5